\documentclass[12pt]{article}

\usepackage{amsmath,mathrsfs,amsfonts,amssymb,graphics,graphicx,epsfig,color,times,bbm}
\usepackage{appendix}

\newcommand{\me}{\mathrm{e}}
\newcommand{\mi}{\mathrm{i}}
\newcommand{\md}{\mathrm{d}}

\newcommand{\nn}{\mathbbm{N}}
\newcommand{\rr}{\mathbbm{R}}
\newcommand{\id}{\mathbbm{1}}
\renewcommand{\vec}[1]{\text{\boldmath$#1$}}

\newtheorem{theorem}{Theorem}
\newtheorem{lemma}{Lemma}
\newtheorem{corollary}{Corollary}

\definecolor{theorem-gray}{rgb}{0.98,0.98,0.98}

\begin{document}

\title{Thermalization under randomized local Hamiltonians}

\author{{\large
M. Cramer}\\ \small Institut f\"ur Theoretische Physik, Albert-Einstein
Allee 11, Universit\"at Ulm, Germany}
\date{}
   \maketitle
\begin{abstract}
Recently, there have been significant new insights concerning conditions under which closed systems equilibrate locally.
The question if subsystems thermalize---if the equilibrium state is independent of the initial state---is however much harder to answer in general.
Here, we consider a setting in which thermalization can be addressed:
A quantum quench under a Hamiltonian whose spectrum is fixed and basis is drawn from the Haar measure. 
If the Fourier transform of the spectral density is small, almost all bases lead to local equilibration to the thermal state with infinite temperature. This allows us to show that, under almost all Hamiltonians that are unitarily equivalent to a local Hamiltonian, it takes an algebraically small time for subsystems to thermalize.

\end{abstract}

\tableofcontents

\section{Introduction}
Consider a large closed system suddenly taken out of equilibrium and evolving unitarily in time. Now suppose
access to the system is limited and only a small subsystem may be observed. 
If this subsystem equilibrates and information about the initial state is lost, principles of statistical mechanics emerge locally: Given sufficient time to equilibrate,
one would observe a generic state---e.g., a (generalized) Gibbs state---which does not depend on the fine-grained information
contained in the initial state anymore but, possibly, instead on macroscopic observables such as, e.g., the mean energy. The system under consideration is closed and the dynamics are unitary. 
Ignorance---the full extent of which one ought to acknowledge frankly \cite{jaynes}, arriving at Jaynes' principle---is replaced by limited  access (spatially -- only subsystems may be observed) to the system. There is no need to introduce ignorance -- no need to put probabilities by hand. Instead, generic ensembles emerge locally from unitary evolution.
This link, along with the recent experimental availability of such {\it quantum quench} settings \cite{experiments1,experiments2,experiments3,experiments4,experiments5,experiments6}, has spurred a bit of a renaissance of ideas dating all the way back to von Neumann's quantum ergodic theorem \cite{vN}. 
An incomplete list of theoretical studies of these questions include the eigenstate thermalization hypothesis \cite{eth1,eth2}, solvable systems \cite{solvable1,solvable2,solvable3,solvable4,solvable5,solvable6},
conformal field theoretical settings \cite{conformal}, and numerical work \cite{numerical}. Conditions guaranteeing and ruling out thermalization have also been given recently \cite{nonthermal1,nonthermal2,nonthermal3}.

For several reasons, the question of equilibration after a quench to a random Hamiltonian has attracted quite some attention lately \cite{fernando,acin,these_other_guys}: A random Hamiltonian may serve as a model for a sufficiently complex system (e.g., certain
properties of quantum chaotic systems are well described by random Hamiltonians \cite{these_other_guys,rev1,rev2}). In fact, in Ref.\ \cite{acin}, a relation between complexity (defined as the number of one- and two-qubit gates needed to approximate the unitary diagonalizing the Hamiltonian) and equilibration properties was suggested. 
Additionally, results on random Hamiltonians have implications on the time scale on which random quantum states may be generated \cite{these_other_guys}. 
Furthermore, to the best of our knowledge, no explicit analytic results concerning the equilibrium state and equilibration time are known for non-integrable systems. As we will see, the setting of random Hamiltonians puts us not only in the position to derive explicit bounds, but also allows us to make the equilibrium state and the time scale of equilibration very explicit. It is also a setting in which usual assumptions on the degeneracy of the energy spectrum are not necessary.

In more technical terms, we are concerned with the following setting. The system is initially in the state $|\psi_0\rangle$ (below we will also allow for mixed initial states) and evolves unitarily according to a Hamiltonian $\hat{H}$, $\hat{\varrho}(t)=\me^{-\mi t\hat{H}}|\psi_0\rangle\langle \psi_0|\me^{\mi t\hat{H}}$. 
We distinguish a subsystem $\mathcal{S}$ and consider the state
on $\mathcal{S}$, $\hat{\varrho}_{\mathcal{S}}(t)=\mathrm{tr}_{\mathcal{B}}[\hat{\varrho}(t)]$, where  $\mathcal{B}$ denotes the rest of the system, which takes the role of a ``bath" or ``environment". Now we ask the question wether there is some canonical time-independent state $\hat{\omega}$ which describes the system locally, i.e., wether $\hat{\varrho}_{\mathcal{S}}(t)$ is close to $\hat{\omega}_{\mathcal{S}}:=\mathrm{tr}_{\mathcal{B}}[\hat{\omega}]$, and if so, for which times. That is, we are concerned with the trace distance $\| \hat{\varrho}_{\mathcal{S}}(t)-\hat{\omega}_{\mathcal{S}}\|_{\mathrm{tr}}$, which quantifies the distinguishability of the two states: For any observable $\hat{O}$ on $\mathcal{S}$ one has $|\langle\hat{O}\rangle_{\hat{\varrho}_{\mathcal{S}}(t)}-\langle\hat{O}\rangle_{\hat{\omega}_{\mathcal{S}}}|\le \|\hat{O}\|\| \hat{\varrho}_{\mathcal{S}}(t)-\hat{\omega}_{\mathcal{S}}\|_{\mathrm{tr}}$, where $\|\cdot\|$ denotes the operator norm.
A seminal result is the following \cite{winter} (see also Ref.\ \cite{reimann}). If the energy gaps of the Hamiltonian $\hat{H}$ are non-degenerate, one has
\begin{equation}
\label{winter_bound}
\lim_{T\rightarrow\infty}\frac{1}{T}\int_0^T\!\!\!\md t\,\|\hat{\varrho}_{\mathcal{S}}(t)-\hat{\omega}_{\mathcal{S}}\|_{\mathrm{tr}}\le
d_{\mathcal{S}}\sqrt{\mathrm{tr}[\hat{\omega}^2]},
\end{equation}
where $d_{\mathcal{S}}$ is the dimension of the Hilbert space associated with $\mathcal{S}$ and $\mathrm{tr}[\hat{\omega}^2]$ is the purity of the equilibrium state
\begin{equation}
\label{equi}
\hat{\omega}=\lim_{T\rightarrow\infty}\frac{1}{T}\int_0^T\!\!\!\md t\, \hat{\varrho}(t)=\sum_{\vec{n}} |\psi_{\vec{n}}|^2|v_{\vec{n}}\rangle\langle v_{\vec{n}}|,
\end{equation} 
where $|v_{\vec{n}}\rangle$ and $E_{\vec{n}}$ are eigenstates and corresponding energies of the Hamiltonian, respectively,
and $\psi_{\vec{n}}=\langle v_{\vec{n}}|\psi_0\rangle$. Hence, this result identifies the equilibrium state and establishes that if its purity is small,
the system is locally well described by $\hat{\omega}$ for almost all times. Two questions remain. What is the time-scale of equilibration, i.e., what can be said for finite times $T$? Further, when is $\hat{\omega}_{\mathcal{S}}$ independent of the initial state, i.e., when does the system not only equilibrate but thermalize in this sense? Progress towards the equilibration time scale was made in the recent Ref.\ \cite{short}, in which also the condition on energy gaps was relaxed, arriving at a bound that also involves the purity of the equilibrium state. 
Picking the Hamiltonian $\hat{H}=\hat{U}\mathrm{diag}[\{E_{\vec{n}}\}]\hat{U}^\dagger$ randomly by fixing the spectrum $\{E_{\vec{n}}\}$ and drawing $\hat{U}$ from the Haar measure,
conditions on the energy gaps may also be relaxed and equilibration time scales obtained \cite{fernando,acin}. The question of initial state independence, i.e., thermalization remains open, however. For the randomized setting just described, we will show that thermalization can be addressed and that the system thermalizes to the maximally mixed state $\hat{\omega}_{\mathcal{S}}=\id/d_{\mathcal{S}}$ -- the Gibbs state with infinite temperature. If the spectrum $\{E_{\vec{n}}\}$ is that of a local Hamiltonian, we will show on which time scales to expect thermalization.

\section{Thermalization under random Hamiltonians} We start with some notation. The Hilbert space under consideration is
\begin{equation}
\nonumber
\begin{split}
\mathcal{H}&=\mathrm{span}\{|\vec{n}\rangle=|n_1\cdots n_N\rangle\,;\,n_i=1,\dots,d_i\}
=\mathcal{S}\otimes\mathcal{B}
\end{split}
\end{equation}
and we denote $d_{\mathcal{S}}=\mathrm{dim}[\mathcal{S}]$, $d_{\mathcal{B}}=\mathrm{dim}[\mathcal{B}]$, $d=\mathrm{dim}[\mathcal{H}]=d_{\mathcal{S}}d_{\mathcal{B}}$. This is the Hilbert space of a collection of $N$ subsystems, each with dimension $d_i$ -- e.g., $N$ spin-$1/2$ particles ($d_i=2$). 
Until we consider local Hamiltonians below, all considerations are independent of the geometry of the modelled system. That is,
we need not assume a specific arrangement of the $N$ subsystems or {\em sites} (such as, e.g., spins on some sort of lattice). Let $\hat{H}$ a Hamiltonian,
\begin{equation}
\label{ham}
\hat{H}=\hat{U}\Bigl(\sum_{\vec{n}}E_{\vec{n}}|\vec{n}\rangle\langle \vec{n}|\Bigr)\hat{U}^\dagger,
\end{equation}
$\hat{\varrho}(t)$ a time-evolved state,
\begin{equation}
\hat{\varrho}(t)=\me^{-\mi t\hat{H}}\hat{\varrho}_0\me^{\mi t\hat{H}},
\end{equation}
and denote $\hat{\varrho}_{\mathcal{S}}(t)=\mathrm{tr}_{\mathcal{B}}[\hat{\varrho}(t)]$. We do not restrict ourselves to pure initial states but
allow $\hat{\varrho}_0$ to be mixed.
We will pick Hamiltonians as in Eq.\ (\ref{ham}) at random by fixing the spectrum $\{E_{\vec{n}}\}$ and sampling $\hat{U}$ from the Haar measure.
We denote
\begin{equation}
\mathbbm{E}[\bullet]=\int\!\bullet\;\md\mu(\hat{U}),
\end{equation}
where $\mu$ is the Haar measure.
We set out to bound the expected trace distance of the local time-evolved state $\hat{\varrho}_{\mathcal{S}}(t)$
to the maximally mixed state. If this distance is small, the system is locally well described by the thermal state with infinite temperature.
 In Ref.\ \cite{these_other_guys} it was shown that for initial pure product states, $\hat{\varrho}_0=|\psi_{\mathcal{S}}\rangle\langle\psi_{\mathcal{S}} |\otimes |\psi_{\mathcal{B}}\rangle\langle \psi_{\mathcal{B}}|$, the expected purity of $\hat{\varrho}_{\mathcal{S}}(t)$, $\mathbbm{E}\bigl[\mathrm{tr}_{\mathcal{S}}[\hat{\varrho}_{\mathcal{S}}^2(t)]\bigr]$, is given by \cite{haar_comment}
\begin{equation}
\label{purity}
\begin{split}
\frac{\delta}{1+d}+\frac{4(d-\delta+1)d^2}{(d+3)(d^2-1)}
\Bigl(\frac{|\phi(t)|^4}{4}+\frac{|\phi(2t)|^2}{4d^2}
+\frac{\Re[ \phi^2(t)\phi^*(2t)]}{2d}-\frac{|\phi(t)|^2}{d^2}\Bigr),
\end{split}
\end{equation}
where $\delta=d_{\mathcal{B}}+d_{\mathcal{S}}$ and
\begin{equation}
\phi(t)=\frac{\mathrm{tr}[\me^{\mi t\hat{H}}]}{d}=\frac{1}{d}\sum_{\vec{k}}\me^{\mi t E_{\vec{k}}}
\end{equation}
is
the Fourier transform of the spectral density. For general states $\hat{\varrho}$, the quantity $\phi_{\hat{\varrho}}(t)=\mathrm{tr}[\me^{\mi t\hat{H}}\hat{\varrho}]$ is also known as {\em characteristic function} (it is a positive definite function and hence, due to Bochner's theorem, the characteristic function of a random variable), taking centre stage in many proofs of quantum central limit theorems (see, e.g., Refs.\cite{goderis_vets,hartmann}), and  $\phi_{|\psi\rangle\langle \psi|}(t)$ is the {\em Loschmidt echo} of $|\psi\rangle$. Statistics and equilibration time of the latter was studied recently in \cite{zanardi}.

As $|\phi(t)|\le 1$, it follows for separable initial states $\hat{\varrho}_0=\sum_np_n\hat{\varrho}^{(n)}_{\mathcal{S}}\!\!\otimes\! \hat{\varrho}^{(n)}_{\mathcal{B}}$ that \cite{sep_note}
\begin{equation}
\nonumber
\begin{split}
\mathbbm{E}\bigl[\mathrm{tr}_{\mathcal{S}}[\hat{\varrho}_{\mathcal{S}}^2(t)]\bigr]-\frac{1}{d_{\mathcal{S}}}&\le
|\phi(t)|^4+ \frac{4}{d_{\mathcal{B}}}
\end{split}
\end{equation}
and we may formulate the following direct consequence of (\ref{purity}), which should be compared to Refs.\ \cite{fernando,acin}, in which a similar bound was given for the expected distance of $\hat{\varrho}_{\mathcal{S}}(t)$ to the (there unknown) equilibrium state $\hat{\omega}_{\mathcal{S}}$ in Eq.\ (\ref{equi}).
\begin{corollary}
Let $\{E_{\vec{n}}\}$ be given, $\hat{\varrho}_0$ a separable state, and $\hat{H}$ as in Eq.\ (\ref{ham}). Then
\begin{equation}
\label{bb}
\mathbbm{E}\bigl[\|\hat{\varrho}_{\mathcal{S}}(t)-\id_{\mathcal{S}}/d_{\mathcal{S}}\|_{\mathrm{tr}}\bigr]\le
\sqrt{d_{\mathcal{S}}}
\sqrt{|\phi(t)|^4+ 4/d_{\mathcal{B}}}.
\end{equation}
\end{corollary}
Hence, if the right hand side is small, we expect the system to be close to the maximally mixed state -- the thermal state with infinite temperature.

We now turn to investigating $\phi(t)$ for two classes of systems for which the above bound can be made explicit and also the thermalization time scale be given explicitly: Solvable systems and general local Hamiltonians.
The latter constitutes the main result of this work while the former illustrates---at the hand of a rather simple proof---what to expect for more general systems. Due to it being rather technical and long, the proof for general local Hamiltonians may be found in the Appendix.

\section{Time scale of thermalization} We write $\langle\cdot\rangle=\mathrm{tr}[\frac{\id}{d}\,\cdot]$ and
\begin{equation}
\sigma^2=\bigl\langle\bigl(\hat{H}-\langle\hat{H}\rangle\bigr)^2\bigr\rangle,\;\;\; \bar{\sigma}^2=\frac{\sigma^2}{N},
\end{equation} 
where we note that for a large class of Hamiltonians $\bar{\sigma}$ is simply a constant \cite{sigma}. 
Further, we denote the time average of the trace distance as 
\begin{equation}
\Delta(T)=\frac{1}{T}\int_0^T\!\!\md t\, \|\hat{\varrho}_{\mathcal{S}}(t)-\id_{\mathcal{S}}/d_{\mathcal{S}}\|_{\mathrm{tr}}.
\end{equation}
We will now give explicit bounds on $\mathbbm{E}[\Delta(T)]$ in terms of $\bar{\sigma}$ and the system size, enabling us to extract the time scale of thermalization. Suppose we obtain the bound $\mathbbm{E}[\Delta(T)]\le c$. We may then also give a bound on the probability that $\Delta(T)\le y c$: From Markov's inequality, we have for all $y>0$ that
\begin{equation}
\mathbbm{P}\bigl[\Delta(T)\le y c\bigr]\ge 1-1/y.
\end{equation}
Further, the fraction of times in $[0,T]$ for which $\|\hat{\varrho}_{\mathcal{S}}(t)-\id_{\mathcal{S}}/d_{\mathcal{S}}\|_{\mathrm{tr}}\le 1/x$ is at least
$1-x\Delta(T)$.
Combining all the above, a bound $\mathbbm{E}[\Delta(T)]\le c$ hence enables us to conclude that with probability at least $1-1/y$, the fraction of times in $[0,T]$ for which $\|\hat{\varrho}_{\mathcal{S}}(t)-\id_{\mathcal{S}}/d_{\mathcal{S}}\|_{\mathrm{tr}}\le 1/x$ is at least $1-xyc$. Hence, choosing $x,y\gg 1$  and supposing $c$ is such that $xyc\ll 1$, we have that for almost all $\hat{U}$ and almost all $t\in [0,T]$, the subsystem is close to the maximally mixed state. We proceed by giving bounds for solvable system and then show that similar bounds hold for spectra of local Hamiltonians.

\subsection{Solvable systems}Assume that the spectrum takes the form
\begin{equation}
\label{spec}
E_{\vec{n}}=\sum_{k=1}^N\epsilon_kn_k,\;\; n_k=0,1.
\end{equation}
This includes, e.g., spin chains solvable via a Jordan-Wigner transformation, i.e., in particular  all Hamiltonians of the form
\begin{equation}
\begin{split}
\hat{H}&=\sum_{i=1}^{N-1}\sum_{\alpha,\beta=x,y}J_{i,\alpha,\beta}\hat{\sigma}_i^\alpha\hat{\sigma}_{i+1}^\beta-\sum_{i=1}^Nh_i\hat{\sigma}_i^z,
\end{split}
\end{equation}
i.e., all $XY$-type models in transverse fields. Straightforward calculations show that
\begin{equation}
\sigma^2=\frac{1}{4}\sum_k\epsilon_k^2
\end{equation}
and
\begin{equation}
|\phi(t)|=\prod_k|\cos(\epsilon_kt/2)|.
\end{equation}
Hence, denoting $\epsilon_{\mathrm{max}}=\max_k|\epsilon_k|$, we have for
 $|t|\epsilon_{\mathrm{max}}\le 2\pi$ that
 \begin{equation}
 |\phi(t)|\le \prod_k\me^{-\epsilon^2_kt^2/8}=\me^{-\sigma^2t^2/2},
 \end{equation}
 which implies that there is an absolute constant $a_0$ such that for $T\epsilon_{\mathrm{max}}\le 2\pi$, we have
\begin{equation}
\mathbbm{E}\bigl[\Delta(T)\bigr]^2\le a_0d_{\mathcal{S}}
\left(\frac{1}{T\sigma}+\frac{1}{d_{\mathcal{B}}}\right).
\end{equation}
In particular, 
\begin{equation}
\mathbbm{E}\bigl[\Delta(T)\bigr]^2\le a_0d_{\mathcal{S}}
\left(\frac{\epsilon_{\mathrm{max}}}{N^{\epsilon}\bar{\sigma}}+\frac{1}{d_{\mathcal{B}}}\right)
\end{equation}
for $T\epsilon_{\mathrm{max}}=N^{\epsilon-1/2}$ and all $0<\epsilon\le 1/2$. Hence, for sufficiently large system size
and almost all Hamiltonians as above, the subsystem will spend most of its time in $[0,T]$, $T\sim N^{\epsilon-1/2}$, close to the maximally mixed state -- under almost all Hamiltonians with a spectrum as in Eq.\ (\ref{spec}), subsystems thermalize in a time $T\sim N^{\epsilon-1/2}$. We will now see that a similar result holds for spectra of local Hamiltonians.

\subsection{Local Hamiltonians}We now consider local Hamiltonians on a $D$-dimensional cubic lattice, which we denote as the collection of sites $\mathcal{L}=\{1,\dots,M\}^{\times D}$, i.e., $N=M^D$. We equip the lattice with a distance $dist(i,j)$, which we take as the shortest path connecting $i$ and $j$. For open boundary conditions, this is simply $dist(i,j)=\sum_{\delta=1}^D|i_\delta-j_\delta|$. The Hamiltonian is assumed to be local in the sense that
\begin{equation}
\hat{H}=\sum_{i\in\mathcal{L}}\hat{h}_i,
\end{equation}
where $\hat{h}_i$ acts non-trivially only on $\{j\in\mathcal{L}\,|\,dist(i,j)\le R\}$ for some constant $0<R\in\nn$. Further, we assume that $\|\hat{h}_i\|\le h$ for all $i$ and some constant $h$ and that the lattice is sufficiently large such that $0<4^{3/2}R\le M^{3/5}$. Hamiltonians as just described, we will simply call local Hamiltonians.
A proof of the following theorem can be found in the appendix. The proof is based on ideas (which we generalized to account for finite system sizes $N$) employed in proofs of quantum central limit theorems \cite{goderis_vets,hartmann}.
 
\begin{theorem} Let $\hat{H}$ a local Hamiltonian. Then there exist constants $a_0$ and $b_0$, only depending on the interaction radius $R$, the interaction strength $h$ and the dimension of the lattice $D$, such that
for 
\begin{equation}
T= a_0 \bar{\sigma}^2N^{1/(5D)-1/2}
\end{equation}
we have
\begin{equation}
\frac{1}{T}\int_0^T\!\!\md t\, |\phi(t)|\le \frac{ b_0 \left(1+\bar{\sigma}^{-3}\right)}{N^{1/(5D)}}.
\end{equation}
\end{theorem}
This together with (\ref{bb}) allows us to formulate our main result.
\begin{corollary} Let $\{E_{\vec{n}}\}$ be the spectrum of a local Hamiltonian. Let $\hat{H}$ as in Eq.\ (\ref{ham}) and $\hat{\varrho}_0$ a separable state. Then there are constants $a_0$ and $b_0$, only depending on the interaction radius $R$, the interaction strength $h$ and the dimension of the lattice $D$, such that
for 
\begin{equation}
\label{T}
T= a_0 \bar{\sigma}^2N^{1/(5D)-1/2}
\end{equation}
we have
\begin{equation}
\nonumber
\begin{split}
\mathbbm{E}\bigl[\Delta(T)\bigr]^2&\le
b_0d_{\mathcal{S}}\left(\frac{1+\bar{\sigma}^{-3}}{N^{1/(5D)}}
+\frac{1}{d_{\mathcal{B}}}\right).
\end{split}
\end{equation}
\end{corollary}
Hence, under almost all Hamiltonians that are unitarily equivalent to a local Hamiltonian, the subsystem will, for sufficiently large system size, spend most of its time in $[0,T]$, $T\sim N^{1/(5D)-1/2}$, close to the maximally mixed state -- under almost all Hamiltonians that are unitarily equivalent to a local Hamiltonian, subsystems thermalize in a time $T\sim N^{1/(5D)-1/2}$. More precisely: Let $\epsilon>0$ and $T$ as in Eq.\ (\ref{T}). Then, with probability at least 
\begin{equation}
1-N^{(\epsilon-\frac{1}{5D})/4},
\end{equation}
the fraction of times in $[0,T]$ for which 
\begin{equation}
\|\hat{\varrho}_{\mathcal{S}}(t)-\id_{\mathcal{S}}/d_{\mathcal{S}}\|_{\mathrm{tr}}\le N^{(\epsilon-\frac{1}{5D})/4}
\end{equation}
is at least
\begin{equation}
1-\sqrt{b_0d_{\mathcal{S}}\left(\frac{1+\bar{\sigma}^{-3}}{N^{\epsilon}}
+\frac{N^{1/(5D)}}{N^{\epsilon}d_{\mathcal{B}}}\right)}.
\end{equation}
Maybe surprisingly, the size of the subsystem does not need to be constant. The above bounds allow for subsystems whose size increases logarithmically with the system size $N$.

\section{Discussion}We have shown on which time scales subsystems thermalize under unitary dynamics generated by randomized local Hamiltonians. 
In this setting, usual assumptions on the degeneracy of energy gaps are not necessary and the equilibrium state---here the maximally mixed state---and bounds can be given explicitly. The only remaining assumption is that the energy variance $\sigma^2$ is lower bounded by the system size, which is also necessary for asymptotic normality \cite{goderis_vets,hartmann} and fulfilled for a large class of Hamiltonians \cite{sigma}. As the system locally equilibrates to the maximally mixed state---the thermal state with infinite temperature---the question arises under which conditions a thermal state with finite temperature might emerge. It would be interesting to see wether restricting the unitaries to the ones preserving the mean energy of the initial state lead to a finite temperature Gibbs state.

\section*{Acknowledgements}
\addcontentsline{toc}{section}{Acknowledgements}
We gratefully acknowledge discussions with Fernando G.S.L.\ Brand\~{a}o. This work was supported by the Alexander von Humboldt Foundation and the
EU STREP projects HIP and PICC.


\appendix
\section{Proof of Theorem 1}
\subsection{Preliminaries}
We denote
$\langle\cdot\rangle=\text{tr}[\cdot]/d$.
For subsets $\mathcal{A}\subset\mathcal{L}$, we write
\begin{equation}
\nonumber
\hat{H}_{\mathcal{A}}=\sum_{i\in\mathcal{A}}\hat{h}_i,\;\;
\sigma_{\mathcal{A}}^2=\bigl\langle\bigl(\hat{H}_{\mathcal{A}}-\langle\hat{H}_{\mathcal{A}}\rangle\bigr)^2\bigr\rangle,\;\;
\phi_{\mathcal{A}}(t)=\langle\me^{\mi t\hat{H}_{\mathcal{A}}}\rangle,
\end{equation}
and omit the index for $\mathcal{A}=\mathcal{L}$. 
We will first prove the following lemma and then make the partition explicit in section \ref{put_together}, which proves theorem 1.

\begin{lemma}Let $\mathcal{L}=\mathcal{A}\cup\mathcal{C}$, $\mathcal{A}=\cup_{n=1}^A\mathcal{A}_n$ a partition of the lattice such that $dist(\mathcal{A}_n,\mathcal{A}_m)>2R$ for $n\ne m$. Then
there are constants $a_0$ and $b_0$, only depending on the interaction radius $R$, the interaction strength $h$ and the dimension of the lattice $D$, such that
for 
\begin{equation}
T\le\tfrac{a_0\sigma^2}{|\mathcal{A}|\max_n|\mathcal{A}_n|^{1/2}}
\end{equation}
we have
\begin{equation}
\tfrac{1}{T}\int_0^T\!\!\!\md t\,|\phi(t)|\le b_0\left( |\mathcal{C}|T^2+\tfrac{1}{T\sigma}\right).
\end{equation}
\end{lemma}
If the lemma holds for $\langle\hat{h}_i\rangle=0$ then, by noting that
\begin{equation}
\begin{split}
|\phi(t)|=|\phi(t)\me^{-\mi t\sum_{i\in\mathcal{L}}\langle\hat{h}_i\rangle}|,
\end{split}
\end{equation}
it holds also for $\langle\hat{h}_i\rangle\ne 0$, i.e., we may assume that $ \langle\hat{h}_i\rangle=0$. Now,
\begin{equation}
\begin{split}
|\phi(t)|&\le |\phi(t)-\phi_{\mathcal{A}}(t)|+\left|\phi_{\mathcal{A}}(t)-\me^{-\sigma^2t^2/2}\right|+\me^{-\sigma^2t^2/2}\\
&=:\delta_1(t)+\delta_2(t)+\me^{-\sigma^2t^2/2}.
\end{split}
\end{equation}
We proceed by bounding $\delta_1(t)$ and $\delta_2(t)$.
The proof techniques are similar to the ones in Ref.\ \cite{goderis_vets} (see also Ref.\ \cite{hartmann}), we however improve on some of the bounds. Several times we
will make use of
\begin{equation}
\label{ball}
\max_{j\in\mathcal{L}}\sum_{\substack{i\in\mathcal{L}\\ dist(i,j)\le r}}1\le 1+c_Dr^D=:\beta_r,
\end{equation}
the constant $c_D$ depending only on the dimension of the lattice.
\subsection{Bound on $\delta_1$}
Denote $\hat{H}_{\mathcal{C}}=\sum_{n=1}^{|\mathcal{C}|}\hat{h}_n$, i.e.,
we slightly abuse notation by using $i\in\mathcal{C}$ and $n=1,\dots,|\mathcal{C}|$ interchangeably.
Writing $\hat{H}_m=\hat{H}-\sum_{n=1}^m\hat{h}_n$, $\hat{H}_0=\hat{H}$, we have
\begin{equation}
\begin{split}
\me^{\mi t\hat{H}}-\me^{\mi t\hat{H}_{\mathcal{A}}} &=\sum_{m=1}^{|\mathcal{C}|}\bigl(\me^{\mi t\hat{H}_{m-1}}-\me^{\mi t\hat{H}_{m}}\bigr)\\
&=\sum_{m=1}^{|\mathcal{C}|}\bigl(\me^{\mi t(\hat{H}_{m}+\hat{h}_m)}\me^{-\mi t\hat{H}_{m}}-\id\bigr)\me^{\mi t\hat{H}_{m}}\\
&=:\sum_{m=1}^{|\mathcal{C}|}\hat{X}_m(t)\me^{\mi t\hat{H}_{m}},
\end{split}
\end{equation}
where
\begin{equation}
\hat{X}_m(t)=\hat{X}_m(0)+t\tfrac{\md\hat{X}_m}{\md t}(0)+\int_0^t\!\!\md r\!\!\int_0^{r}\!\!\md s\,\tfrac{\md^2 \hat{X}_m}{\md t^2}(s),
\end{equation}
$\hat{X}_m(0)=0$, $\tfrac{\md\hat{X}_m}{\md t}(0)=\mi \hat{h}_m$, and the last term is equal to
\begin{equation}
\int_0^t\!\!\md r\!\!\int_0^{r}\!\!\md s\,\me^{\mi s\hat{H}_{m-1}}\bigl([\hat{h}_m,\hat{H}_m]-\hat{h}_m^2\bigr)\me^{-\mi s\hat{H}_m},
\end{equation}
i.e.,
\begin{equation}
\nonumber
\begin{split}
|\langle\hat{X}_m\me^{\mi t\hat{H}_{m}}\rangle|\le
|t|
|\langle \hat{h}_m\me^{\mi t\hat{H}_{m}}\rangle|
+\tfrac{t^2}{2}\|[\hat{h}_m,\hat{H}_m]\|
+\tfrac{t^2h^2}{2},
\end{split}
\end{equation}
where (recall that $\hat{h}_i$ and $\hat{h}_j$ commute for $dist(i,j)>2R$)
\begin{equation}
\nonumber
\begin{split}
\|[\hat{h}_m,\hat{H}_m]\|&\le \sum_{\substack{i\in\mathcal{L} \\ dist(i,m)\le 2R }}\|[\hat{h}_m,\hat{h}_i]\|\le h^2 \beta_{2R}.
\end{split}
\end{equation}
Now, $\hat{H}_{m}=\sum_{i\in\mathcal{X}_m}\hat{h}_i$ for some $\mathcal{X}_m\subset\mathcal{L}$, i.e., writing 
\begin{equation}
\hat{H}_{m}=\sum_{\substack{i\in\mathcal{X}_m \\ dist(i,m)\le 2R}}\hat{h}_i+\sum_{\substack{i\in\mathcal{X}_m \\ dist(i,m)> 2R}}\hat{h}_i
=:\hat{Y}_m+\hat{Z}_m,
\end{equation}
we have $\langle\hat{h}_m\me^{\mi t\hat{Z}_{m}}\rangle=0$, i.e.,
\begin{equation}
\begin{split}
\nonumber
|\langle\hat{h}_m\me^{\mi t\hat{H}_{m}}\rangle|&=
|\langle\hat{h}_m\bigl(\me^{\mi t(\hat{Y}_{m}+\hat{Z}_{m})}\me^{-\mi t\hat{Z}_{m}}-\id\bigr)\me^{\mi t\hat{Z}_{m}}\rangle|\\
&\le h\|\me^{\mi t(\hat{Y}_{m}+\hat{Z}_{m})}\me^{-\mi t\hat{Z}_{m}}-\id\|\\
&\le h\int_0^t\!\!\!\md s\,\|\hat{Y}_{m}\|
\le h^2\beta_{2R}|t|.
\end{split}
\end{equation}
Therefore, $\delta_1(t)\le
|\mathcal{C}|\tfrac{h^2t^2}{2}\left(1+3\beta_{2R}\right)$.

\subsection{Bound on $\delta_2$}
Writing
\begin{equation}
\phi_n(t)=\langle\me^{\mi t\hat{H}_n}\rangle,\;\;\; \hat{H}_n=\sum_{i\in\mathcal{A}_n}\hat{h}_i,
\end{equation}
we have
\begin{equation}
\nonumber
\delta_2(t)=\me^{-\sigma^2t^2/2}\Bigl|\me^{\sigma^2t^2/2}\prod_{n=1}^A\phi_n(t)-1\Bigr|.
\end{equation}
Taylor's theorem implies
\begin{equation}
|\phi_n(t)-1|\le\langle\hat{H}^2_n\rangle\tfrac{t^2}{2}=:\tfrac{\sigma_n^2t^2}{2},
\end{equation}
i.e., if $\sigma_n^2t^2<2$, we may take a logarithm and, noting that $\sigma_{\mathcal{A}}^2=\sum_{n=1}^A\sigma_n^2$, write
\begin{equation}
\nonumber
\begin{split}
\delta_2(t)&=\me^{-\sigma^2t^2/2}\left|\me^{(\sigma^2-\sigma_{\mathcal{A}}^2)t^2/2}\me^{\sum_{n=1}^A(\sigma_{n}^2t^2/2+\log[\phi_n(t)])}-1\right|\\
&\le \me^{(|\sigma^2-\sigma_{\mathcal{A}}^2|-\sigma^2)t^2/2}\me^{F(t)}
\left(\tfrac{|\sigma^2-\sigma_{\mathcal{A}}^2|t^2}{2}+F(t)\right),
\end{split}
\end{equation}
where $F(t)=\sum_{n=1}^Af_n(t)$,
\begin{equation}
\nonumber
\begin{split}
f_n(t)&=|\tfrac{\sigma_{n}^2t^2}{2}+\log[\phi_n(t)]|\\
&\le |\phi_n(t)-1+\tfrac{\sigma_{n}^2t^2}{2}|+|\phi_n(t)-1-\log[\phi_n(t)]|\\
&\le \langle|\hat{H}_n|^3\rangle\tfrac{|t|^3}{6}+\tfrac{\sigma_n^3|t|^3}{2^{5/2}}\\
&\le \sqrt{\langle \hat{H}_n^4\rangle\sigma_n^2}\tfrac{|t|^3}{6}+\tfrac{\sigma_n^2t^2}{4}
\end{split}
\end{equation}
where we used Taylor's theorem, Cauchy-Schwarz, and that for all $|z-1|<1$ one has $2|z-\log(1+z)|\le |z|^{3/2}$. Now, $\langle \hat{H}_n^4\rangle\le c_4|\mathcal{A}_n|^2$ and $\sigma_{n}^2\le c_2|\mathcal{A}_n|$ (see Section \ref{moment_bounds} below). Hence,
\begin{equation}
-\tfrac{\sigma_{n}^2t^2}{2}+f_n(t)\le \sqrt{c_2c_4}|\mathcal{A}_n|^{3/2}\tfrac{|t|^3}{6}-\tfrac{\sigma_n^2t^2}{4}.
\end{equation}
For all $T>0$, we have
\begin{equation}
|\mathcal{C}|T^2+\tfrac{1}{T\sigma}\ge \bigl(\tfrac{27|\mathcal{C}|}{4\sigma^{2}}\bigr)^{1/3},
\end{equation}
i.e., for $81 c_2|\mathcal{C}|\ge 4\sigma^{2}$, the lemma is trivially true:
\begin{equation}
\nonumber
\tfrac{1}{T}\int_0^T\!\!\!\md t\,|\phi(t)|\le 1 \le \bigl(\tfrac{81c_2|\mathcal{C}|}{4\sigma^{2}}\bigr)^{1/3}\le (3c_2)^{1/3} |\mathcal{C}|T^2+\tfrac{(3c_2)^{1/3}}{T\sigma}
\end{equation}
for all $T>0$. We may hence assume that $81 c_2|\mathcal{C}|\le 4\sigma^{2}$. As $|\sigma^{2}-\sigma_{\mathcal{A}}^{2}|\le 3c_2|\mathcal{C}|$ (see Section \ref{moment_bounds} below), we hence have $23\sigma^{2}\le 27\sigma_{\mathcal{A}}^{2}$. Now, for $x>0$
\begin{equation}
\label{tbound}
|t|\le \tfrac{x\sigma^2}{|\mathcal{A}|\max_n|\mathcal{A}_n|^{1/2}}
\end{equation}
implies
\begin{equation}
t^2\sigma_n^2\le c_2|\mathcal{A}_n|t^2\le \tfrac{c_2|\mathcal{A}_n|x^2\sigma^4_{\mathcal{A}}}{|\mathcal{A}|^2\max_n|\mathcal{A}_n|}
\le c^3_2x^2
\end{equation}
and
\begin{equation}
\nonumber
\begin{split}
\tfrac{(|\sigma^2-\sigma_{\mathcal{A}}^2|-\sigma^2)t^2}{2}+F(t)&\le -\tfrac{23\sigma^2t^2}{54}+\sqrt{c_2c_4}x\sigma^2\tfrac{t^2}{6}-\tfrac{\sigma_{\mathcal{A}}^2t^2}{4}\\
&\le
-\tfrac{\sigma^2t^2}{6}(\tfrac{23}{6}-\sqrt{c_2c_4}x).
\end{split}
\end{equation}
Hence, setting $x=\min\{\sqrt{1/c_2^3},22/(6\sqrt{c_2c_4})\}$ implies $\sigma_n^2t^2<2$ and
\begin{equation}
\tfrac{(|\sigma^2-\sigma_{\mathcal{A}}^2|-\sigma^2)t^2}{2}+F(t)\le
-\tfrac{\sigma^2t^2}{36}.
\end{equation}
Thus, for $t$ as in Eq.\ (\ref{tbound}), we have
\begin{equation}
\nonumber
\begin{split}
\delta_2(t)&\le \me^{-\sigma^2t^2/36}
\left(\tfrac{|\sigma^2-\sigma_{\mathcal{A}}^2|t^2}{2}+F(t)\right)\le \tfrac{17\sigma^2}{36}t^2\me^{-\sigma^2t^2/36}.
\end{split}
\end{equation}

\subsection{Putting it together}
\label{put_together}
So far, we have that there are constants $a_0$ and $b_0$, only depending on $R$, $h$, and $D$, such that for
\begin{equation}
T\le \tfrac{a_0\sigma^2}{|\mathcal{A}|\max_n|\mathcal{A}_n|^{1/2}}
\end{equation}
we have 
\begin{equation}
\nonumber
\begin{split}
\tfrac{1}{Tb_0}\int_0^T\!\!\!\md t\,|\phi(t)|&\le |\mathcal{C}|T^2+\tfrac{1}{T\sigma}\int_0^{\sigma T}\!\!\!\md t\,\left(t^2\me^{-t^2/36}+\me^{-t^2/2}\right)\\
&\le |\mathcal{C}|T^2+\tfrac{1}{T\sigma}\int_0^{\infty}\!\!\!\md t\,(t^2+1)\me^{-t^2/36},
\end{split}
\end{equation}
which proves the lemma. To prove theorem 1, we choose the sets $|\mathcal{A}_n|$ as follows. Let $K=\lfloor \frac{M}{M^{3/5}+2R}\rfloor$, $a=M/K-2R$, and for $k=0,\dots,K-1$, define $\mathcal{X}=\{1,\dots,M\}^{\times (D-1)}$, $a_k=(a+2R)k$,
\begin{equation}
\begin{split}
\mathcal{A}_k&=\mathcal{X}\times\{n\in\nn\,\big|\, a_k< n\le a_k+a\}\\
&=\mathcal{X}\times\{ \lceil a_k-1\rceil,\dots,\lfloor a_k+a\rfloor\},\\
\mathcal{C}_k&=\mathcal{X}\times\{n\in\nn\,\big|\,a_k+a< n \le a_{k+1}\}\\
&=\mathcal{X}\times\{ \lceil a_k+a-1\rceil,\dots,\lfloor a_{k+1}\rfloor\},
\end{split}
\end{equation}
$\mathcal{C}=\bigcup_{k=0}^{K-1}\mathcal{C}_k$. Then, for $n\ne m$,
\begin{equation}
dist(\mathcal{A}_n,\mathcal{A}_m)\ge \lfloor a_{k+1}\rfloor-\lceil a_k+a-1\rceil+2
\ge 2R+1.
\end{equation}
Assuming $M^{3/5}-2R\ge M^{3/5}/2$ and $M-2M^{3/5}\ge M/2$  implies
\begin{equation}
\tfrac{M^{2/5}}{4}\le \tfrac{M-M^{3/5}-2R}{M^{3/5}+2R}\le
K\le M^{2/5}
\end{equation}
and 
\begin{equation}
\begin{split}
\tfrac{M^{D-2/5}}{2}&\le M^{D-1}a=M^{D-1}(M/K-2R)\\
&\le
|\mathcal{A}_k|=M^{D-1}(\lfloor a_k+a\rfloor-\lceil a_k-1\rceil+1)\\
&\le M^{D-1}(M/K-2R+2)\le \tfrac{M^D}{K}\\
&\le 4M^{D-2/5}.
\end{split}
\end{equation}
Further,
\begin{equation}
\begin{split}
|\mathcal{C}_k|&= M^{D-1}(\lfloor a_{k+1}\rfloor-\lceil a_k+a-1\rceil+1)\\
&\le 2( R+1)M^{D-1}.
\end{split}
\end{equation}
Hence, putting
\begin{equation}
T=\tfrac{a_0M^{1/5-D/2}}{8}
\bar{\sigma}^2
\le\tfrac{a_0\sigma^2}{|\mathcal{A}|\max_n|\mathcal{A}_n|^{1/2}},
\end{equation}
we have, redefining $b_0$,
\begin{equation}
\nonumber
\begin{split}
\tfrac{1}{Tb_0}\int_0^T\!\!\!\md t\,|\phi(t)|&\le M^{-1/5}+\tfrac{M^{3D/2-1/5}}{\sigma^3}.
\end{split}
\end{equation}
Finally, the conditions $M-2M^{3/5}\ge M/2$ and $M^{3/5}-2R\ge M^{3/5}/2$ are implied by $0<4^{3/2}R\le M^{3/5}$, which also implies $K\ge 1$, 
$|\mathcal{A}_k|\ge 1$, and $|\mathcal{C}_k|\ge 1$.

\subsection{Some bounds}
\label{moment_bounds}
In this section we will derive the bounds $\sigma_{n}^2\le c_2|\mathcal{A}_n|$, $\langle \hat{H}_n^4\rangle\le c_4|\mathcal{A}_n|^2$, and 
$|\sigma^{2}-\sigma_{\mathcal{A}}^{2}|\le 3c_2|\mathcal{C}|$. We recall that $\langle\hat{h}_i\rangle=0$ for all $i$, that $\langle \hat{h}_i\hat{h}_j\rangle=0$ for $dist(i,j)>2R$, and the bound in Eq.\ (\ref{ball}). We have
\begin{equation}
\begin{split}
\sigma_{n}^2&=\sum_{\substack{i,j\in\mathcal{A}_n \\ dist(i,j)\le 2R}}\langle\hat{h}_i\hat{h}_j\rangle
\le h^2 \sum_{\substack{i,j\in\mathcal{A}_n \\ dist(i,j)\le 2R}}1
\le h^2\beta_{2R}|\mathcal{A}_n|=:c_2|\mathcal{A}_n|.
\end{split}
\end{equation}
Similarly,
\begin{equation}
\begin{split}
\sigma^{2}-\sigma_{\mathcal{A}}^{2}&=\langle \hat{H}_{\mathcal{C}}^2\rangle-2\langle \hat{H}_{\mathcal{A}}\hat{H}_{\mathcal{C}}\rangle
\le c_2|\mathcal{C}|-2\langle \hat{H}_{\mathcal{A}}\hat{H}_{\mathcal{C}}\rangle,
\end{split}
\end{equation}
where
\begin{equation}
\begin{split}
\left|\langle \hat{H}_{\mathcal{A}}\hat{H}_{\mathcal{C}}\rangle\right|&\le\sum_{\substack{i\in\mathcal{A}, j\in\mathcal{C} \\ dist(i,j)\le 2R}}\left|
\langle\hat{h}_i\hat{h}_j\rangle\right|\le h^2\beta_{2R}|\mathcal{C}|=c_2|\mathcal{C}|,
\end{split}
\end{equation}
i.e., $|\sigma^{2}-\sigma_{\mathcal{A}}^{2}|\le 3c_2|\mathcal{C}|$.
To show that $\langle \hat{H}_n^4\rangle\le c_4|\mathcal{A}_n|^2$, we define for given $i\in\mathcal{A}_n$ 
\begin{equation}
\hat{H}_n=\sum_{j\in\mathcal{A}_n}\hat{h}_j=\sum_{\substack{j\in\mathcal{A}_n\\ dist(i,j)\le 2R}}\hat{h}_j+\sum_{\substack{j\in\mathcal{A}_n\\ dist(i,j)> 2R}}\hat{h}_j=:\hat{X}_i+\hat{Y}_i,
\end{equation}
for which we have $[\hat{h}_i,\hat{Y}_i]=0$ and $\langle \hat{h}_i\hat{Y}_i^3\rangle=0$, i.e.,
\begin{equation}
\begin{split}
\langle\hat{H}_{n}^4\bigr\rangle&=\sum_{i\in\mathcal{A}_n}\langle\hat{h}_i(\hat{X}_i^3+\hat{X}_i\hat{Y}_i\hat{X}_i+2\hat{X}_i^2\hat{Y}_i+3\hat{X}_i\hat{Y}^2_i)\rangle\\
&\le 3\sum_{i\in\mathcal{A}_n}\langle\hat{h}_i\hat{X}_i\hat{Y}^2_i\rangle
+h\sum_{i\in\mathcal{A}_n}\left(\|\hat{X}_i\|^3+3\|\hat{X}_i\|^2\|\hat{Y}_i\|\right),
\end{split}
\end{equation}
where we used the cyclic property of the trace.
Now define
\begin{equation}
\hat{Y}_i=\sum_{\substack{j\in\mathcal{A}_n\\ dist(i,j)> 2R}}\hat{h}_j=
\sum_{\substack{j\in\mathcal{A}_n\\ 2R< dist(i,j)\le 4R }}\hat{h}_j
+\sum_{\substack{j\in\mathcal{A}_n\\ dist(i,j)> 4R}}\hat{h}_j
=:\hat{R}_i+\hat{Z}_i
\end{equation}
for which we have $[\hat{h}_i,\hat{Z}_i]=0$, $[\hat{X}_i,\hat{Z}_i]=0$, and $\langle\hat{h}_i\hat{X}_i\hat{Z}^2_i\rangle=0$, i.e., due to the cyclic property of the trace,
\begin{equation}
\begin{split}
\sum_{i\in\mathcal{A}_n}\langle\hat{h}_i\hat{X}_i\hat{Y}^2_i\rangle&=
\sum_{i\in\mathcal{A}_n}\langle\hat{h}_i\hat{X}_i(\hat{R}_i^2+2\hat{R}_i\hat{Z}_i)\rangle\\
&\le h\sum_{i\in\mathcal{A}_n} \left(\|\hat{X}_i\|\|\hat{R}_i\|^2+2\|\hat{X}_i\|\|\hat{R}_i\|\|\hat{Z}_i\|\right).
\end{split}
\end{equation}
Noting that
\begin{equation}
\|\hat{X}_i\|\le  h\beta_{2R},\;\;\;
\|\hat{Y}_i\|\le h|\mathcal{A}_n|,\;\;\;
\|\hat{Z}_i\|\le h|\mathcal{A}_n|,\;\;\;
\|\hat{R}_i\|\le  h\beta_{4R},
\end{equation}
we finally have
\begin{equation}
\begin{split}
\langle\hat{H}_{n}^4\bigr\rangle&\le h^4\left[(3 \beta_{2R}\beta^2_{4R}+\beta^3_{2R})|\mathcal{A}_n|+(6\beta_{2R}\beta_{4R}+3\beta^2_{2R})|\mathcal{A}_n|^2\right]\\
&\le h^4\left[3 \beta_{2R}\beta^2_{4R}+\beta^3_{2R}+6\beta_{2R}\beta_{4R}+3\beta^2_{2R}\right]|\mathcal{A}_n|^2
=:c_4|\mathcal{A}_n|^2.
\end{split}
\end{equation}


\begin{thebibliography}{99}
\bibitem{haar_comment} This involves a very tedious calculation indeed, even without caring about the exact expression for the purity and just arriving at the bound in Eq.\ (\ref{bb}) requires a lot of bookkeeping. After the author did the latter, he learned about Ref.\ \cite{these_other_guys}, which shortened the present manuscript by several pages. The required averages over monomials were derived in Ref.\ \cite{collins} and the necessary Weingarten coefficients may be found in Ref.\ \cite{carlos}.
\bibitem{sep_note} As $\hat{\varrho}_0$ is a convex combination of pure product states, $\hat{\varrho}_0=\sum_mq_m|\psi^{(m)}_{\mathcal{S}}\rangle|\psi^{(m)}_{\mathcal{B}}\rangle\langle\psi^{(m)}_{\mathcal{S}}| \langle\psi^{(m)}_{\mathcal{B}}|=:\sum_mq_m|\psi_{m}\rangle\langle\psi_{m}|$, we have, writing $\hat{\varrho}_{m}(t)=\me^{-\mi t\hat{H}}|\psi_{m}\rangle\langle\psi_{m}|\me^{\mi t\hat{H}}$ and employing the Cauchy-Schwarz inequality, that $\mathrm{tr}_{\mathcal{S}}[\hat{\varrho}_{\mathcal{S}}^2(t)]\le \sum_mq_m\mathrm{tr}_{\mathcal{S}}[\mathrm{tr}_{\mathcal{B}}[\hat{\varrho}_{m}(t)]^2]$.
\bibitem{sigma} Consider, e.g., 
a general translationally invariant nearest neighbour spin Hamiltonian on a chain (generalisations to higher dimensions are straightforward) with periodic boundary conditions, i.e., $\hat{H}=h_0\id+\sum_{i=1}^N\hat{h}_i$
with $
\hat{h}_i=
\sum h_{\alpha,\beta}\hat{\sigma}_i^\alpha\hat{\sigma}_{i+1}^\beta$, $h_{\alpha,\beta}\in\rr$, and summation is over $\alpha=x,y,z$; $\beta=x,y,z,0$.
For Hamiltonians of this form, one finds $\bar{\sigma}^2=\sum h^2_{\alpha,\beta}$.
\end{thebibliography}
\end{document}